# Identifying Autism-Related Neurobiomarkers Using Hybrid Deep Learning Models

Ashley Chen[1]

[1] Oakton High School, 2900 Sutton Rd, Vienna VA 22181, USA

**Abstract.**

**Background:** Autism spectrum disorder (ASD) has been associated with structural alterations across cortical and subcortical regions. Quantitative neuroimaging enables large-scale analysis of these neuroanatomical patterns.

**Methods:** This project used structural MRI (T1-weighted) data from the publicly available ABIDE I dataset (n = 1,112) to classify ASD and control participants using a hybrid model. A 3D convolutional neural network (CNN) was trained to learn neuroanatomical feature representations, which were then passed to a support vector machine (SVM) for final classification. Gradient-weighted class activation mapping (Grad-CAM) was applied to the CNN to visualize the brain regions that contributed most to the model's predictions.

**Results:** Grad-CAM difference maps showed strongest relevance along cortical boundary regions, with additional emphasis in midline frontal-temporal-parietal areas.

**Conclusion:** The hybrid 3D CNN-SVM model classified ASD from ABIDE I structural MRI and highlighted midline frontal-temporal-parietal regions as key contributors, broadly consistent with prior ASD neuroimaging findings.

**Keywords:** autism spectrum disorder, MRI, Gradient-weighted class activation mapping, convolutional neural networks, support vector machines, biomarkers

## 1 Introduction

Autism spectrum disorder (ASD) is a neurodevelopmental condition characterized by differences in social communication and restricted or repetitive behaviors, with frequent sensory processing differences[1]. Clinical identification primarily relies on behavioral assessments such as the Autism Diagnostic Observation Schedule (ADOS-2) and the Autism Diagnostic Interview-Revised (ADI-R). Although widely used, these assessments depend on clinician judgment and context, which can introduce variability across evaluators and settings[2].

The need for complementary, more objective markers has intensified as ASD identification has increased over time. The U.S. Centers for Disease Control and Prevention's (CDC's) Autism and Developmental Disabilities Monitoring (ADDM) Network reports that estimated prevalence rose from about 1 in 150 children in 2000 to about 1 in 36 in 2020[3].

Neuroimaging offers a potential route toward more objective ASD biomarkers because structural MRI can capture neuroanatomical variation across distributed brain systems. Deep learning models such as 3D convolutional neural networks (CNNs) can learn complex spatial patterns directly from volumetric images, offering new opportunities to capture subtle and spatially distributed neuroanatomical differences that may not be evident with traditional analysis. CNNs have shown strong performance in medical imaging, including mammography[4], brain tumor grading[5], and diabetic retinopathy screening[6]. These successes suggest that CNNs are well suited for ASD research, where subtle structural variations across multiple brain regions must be identified.

Additionally, gradient weighted class activation mapping (Grad-CAM) can be used to visualize which brain regions most influenced the CNN's predictions. Grad-CAM computes a class specific importance map by using the gradients of the target class score with respect to the final convolutional feature maps, producing a coarse localization heatmap of the regions that contribute most strongly to that class decision[7]. When applied to 3D structural MRI models, Grad-CAM can be generated as a volumetric relevance map and summarized across participants to create group averaged maps and ASD minus control difference maps, supporting neuroanatomical interpretation of the features learned by the network[8].

Despite their potential, Grad-CAM relevance maps have several limitations. Because Grad-CAM is computed from the final convolutional layers, the resulting heatmaps are inherently low-resolution and tend to provide coarse localization rather than precise anatomical boundaries[9]. The maps are also model-dependent and can vary with architectural choices, the selected layer used for backpropagation, and training variability, which can reduce stability across reruns[10]. More broadly, saliency and attribution methods can sometimes produce visually plausible maps that are not strongly tied to the learned model or can be misleading without validation checks, meaning Grad-CAM should be interpreted as an explanation of the model's decision process rather than a direct map of pathology. Finally, explanation methods can be vulnerable to small input perturbations that substantially alter the explanation while leaving the prediction similar, highlighting the need for robustness and sanity testing when interpreting activation maps.

Even with these limitations, Grad-CAM maps can still provide useful neuroanatomical context by indicating where a trained CNN is extracting class-relevant signals, which supports hypothesis generation and helps relate model behavior to known brain systems. Grad-CAM was designed as a class-discriminative localization method and is often used to identify regions that drive predictions, so consistent group-level patterns across folds, sites, or cohorts can be treated as candidate regions for follow-up with complementary analyses such as atlas-based regional metrics, ROI testing, or multimodal validation[7]. When paired with recommended validation practices such as randomization based sanity checks and robustness assessments, Grad-CAM can also help evaluate whether highlighted structures reflect learned neuroanatomical features rather than spurious cues, supporting more trustworthy interpretation in medical imaging applications[11].

This study trained a 3D CNN–SVM classifier on ABIDE I structural MRI and used Grad-CAM to localize neuroanatomical regions contributing to ASD classification. A 3D CNN was trained on ABIDE I T1-weighted structural MRI to learn feature representations, these features were classified using a support vector machine (SVM), and Grad-CAM was used to visualize the brain regions most associated with ASD versus control predictions. The aim was to identify the neuroanatomical regions associated with autism classification, thereby advancing progress toward objective neuroimaging biomarkers for ASD.

## 2 Methodology

### 2.1 Dataset

This study used structural magnetic resonance imaging (sMRI) data from the Autism Brain Imaging Data Exchange (ABIDE I), a publicly available repository aggregating neuroimaging and clinical data from 17 international research sites, including both individuals diagnosed with autism spectrum disorder (ASD) and typically developing controls (TDC).

In total, 1,112 participants were included, comprising 539 individuals with ASD and 573 controls. Participant ages ranged from childhood to adulthood (mean site ages between ~10 and ~33 years), and the sample reflected the known male predominance in ASD.

Structural MRI provides high-resolution anatomical images of the brain, capturing cortical thickness, white matter, and subcortical morphology. This modality was selected because it offers stable, reproducible measures of brain anatomy suitable for machine learning pipelines.

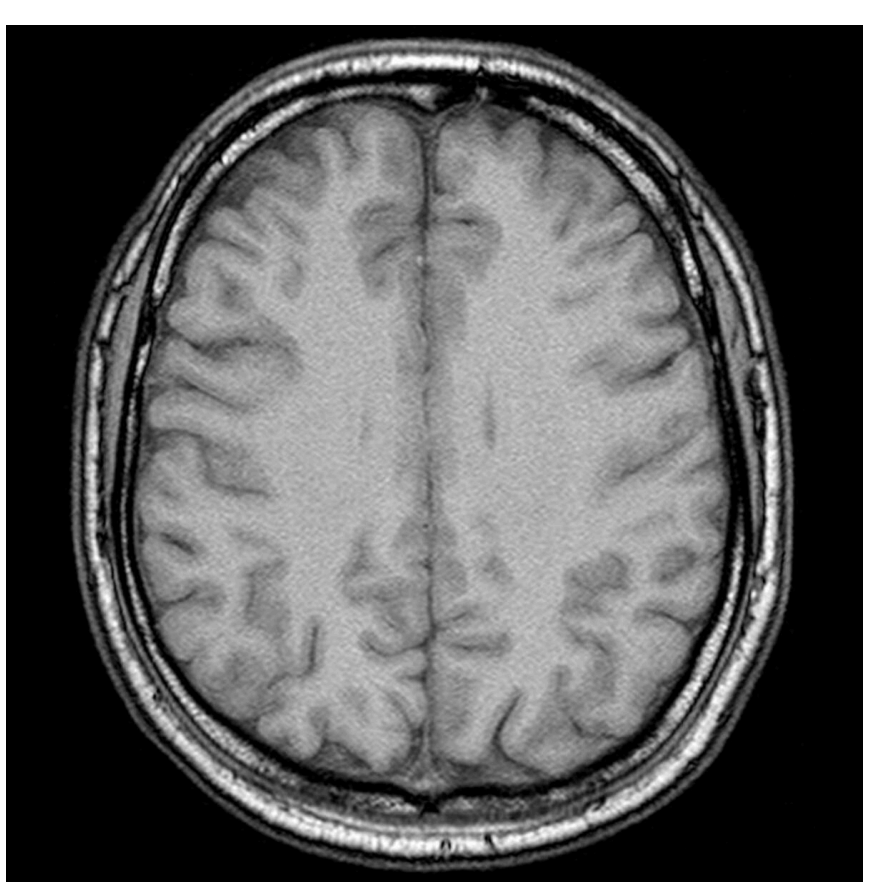

**Fig. 1.** Example of a T1-weighted structural MRI scan from a typically developing control (TDC) participant in the ABIDE I dataset

### 2.2 Preprocessing

Before analysis, raw MRI scans underwent extensive preprocessing to account for heterogeneity across the 17 acquisition sites in ABIDE 1 and to prepare the data for input into convolutional neural networks (CNN). These steps were essential for reducing scanner-related variability while preserving biologically meaningful structural features relevant to autism. The preprocessing pipeline included the following stages:

1. **Skull Stripping.** Non-brain tissue such as scalp, skull, and dura was removed using the HD-BET algorithm, a deep-learning based brain extraction tool. This ensured that subsequent analyses focused exclusively on brain tissue.

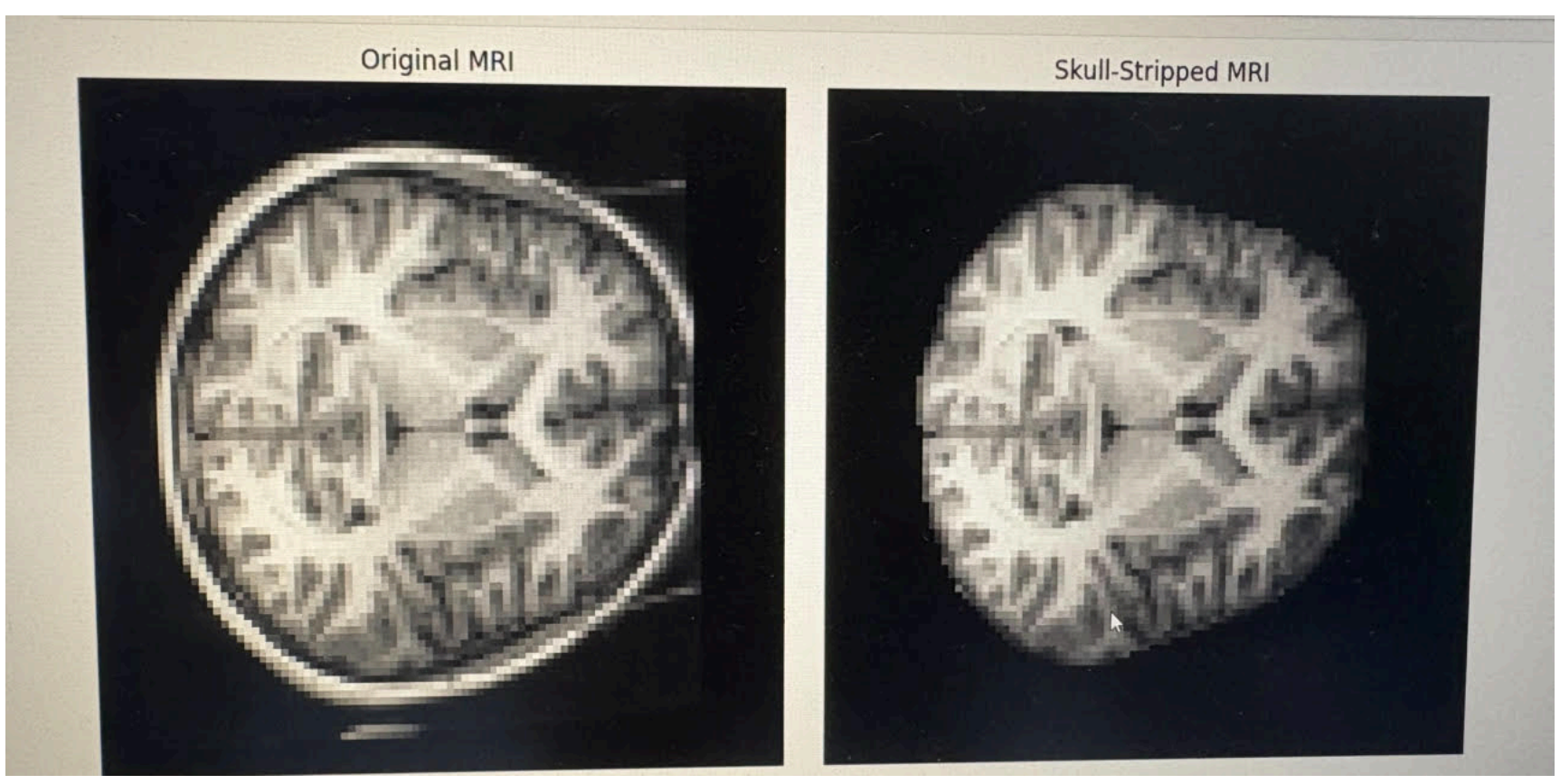

**Fig 2.** Comparison between an original MRI scan (left) and the same scan after skull stripping (right), showing removal of non-brain tissue

2. **Spatial Normalization.** Each scan was registered to the MNI152 standard brain template using Advanced Normalization Tools (ANTs). This step placed all participant images into a common stereotactic space, aligning anatomical structures across subjects and reducing variation due to head position or scanner orientation.

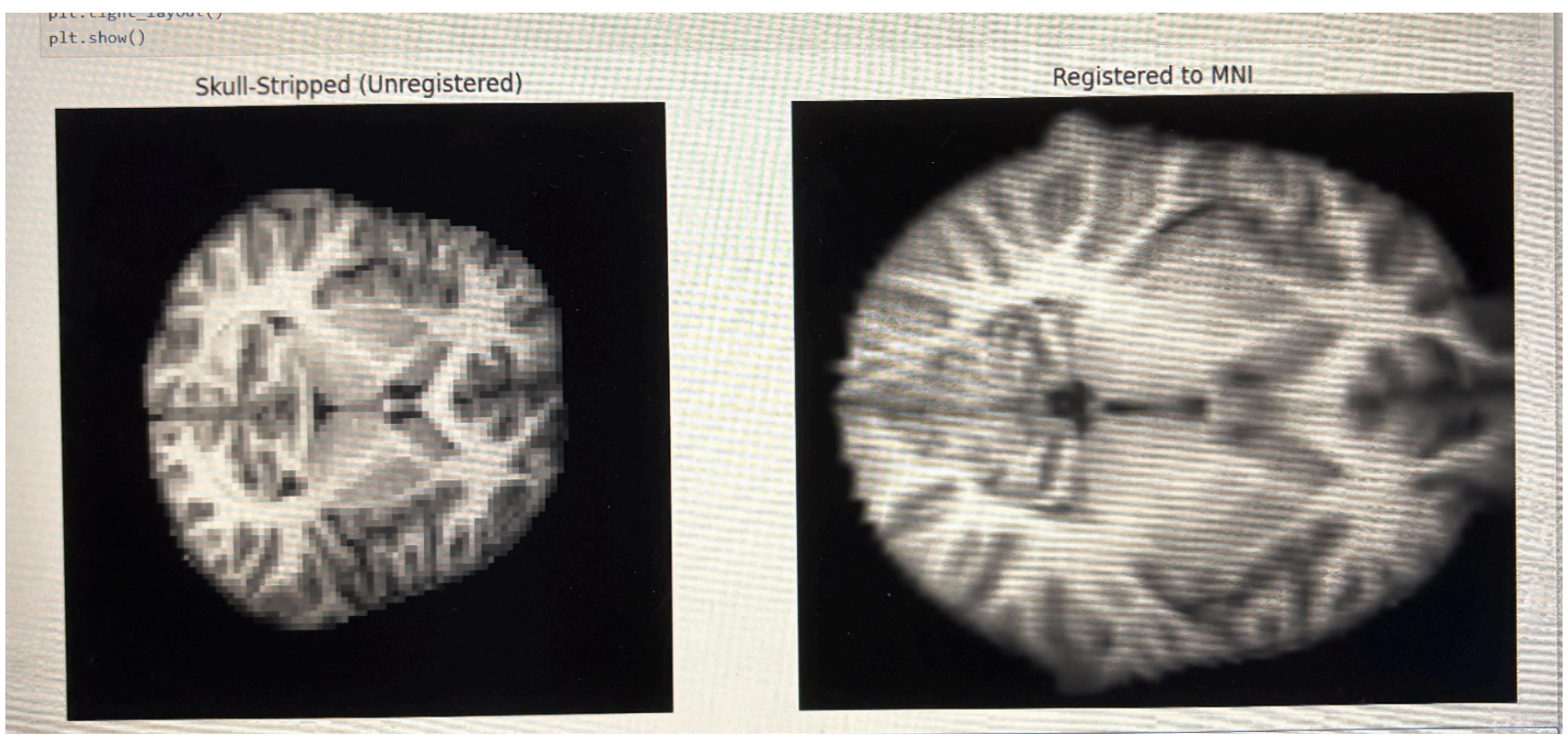

**Fig 3.** MRI scan after skull stripping (left) compared with the same scan registered to the MNI template (right)

3. **Intensity Normalization.** MRI intensity values vary systematically across scanners and acquisition protocols, which can introduce site-specific artifacts into multi-site datasets such as ABIDE. To mitigate these effects, voxel intensities were normalized using a **histogram standardization**. This method aligns the intensity distribution of each subject's scan to a common reference distribution, thereby reducing scanner-related variability while preserving biologically meaningful signal.

   Formally, each voxel intensity $I$ was transformed as:

$$I' = F_{\text{ref}}^{-1}\left(F(I)\right)$$

where *F(I)* is the cumulative distribution function (CDF) of the subject's intensity histogram, and $F_{\mathrm{ref}}^{-1}$ is the inverse CDF of a chosen reference histogram derived using training data only (within each outer fold), then applied unchanged to that fold's validation/test subjects. To prevent leakage, the reference histogram was recomputed separately in each outer fold using only the outer-training partition.

This mapping ensures that intensities are rescaled consistently across sites, preventing the CNN from inadvertently learning scanner differences instead of ASD-related features.

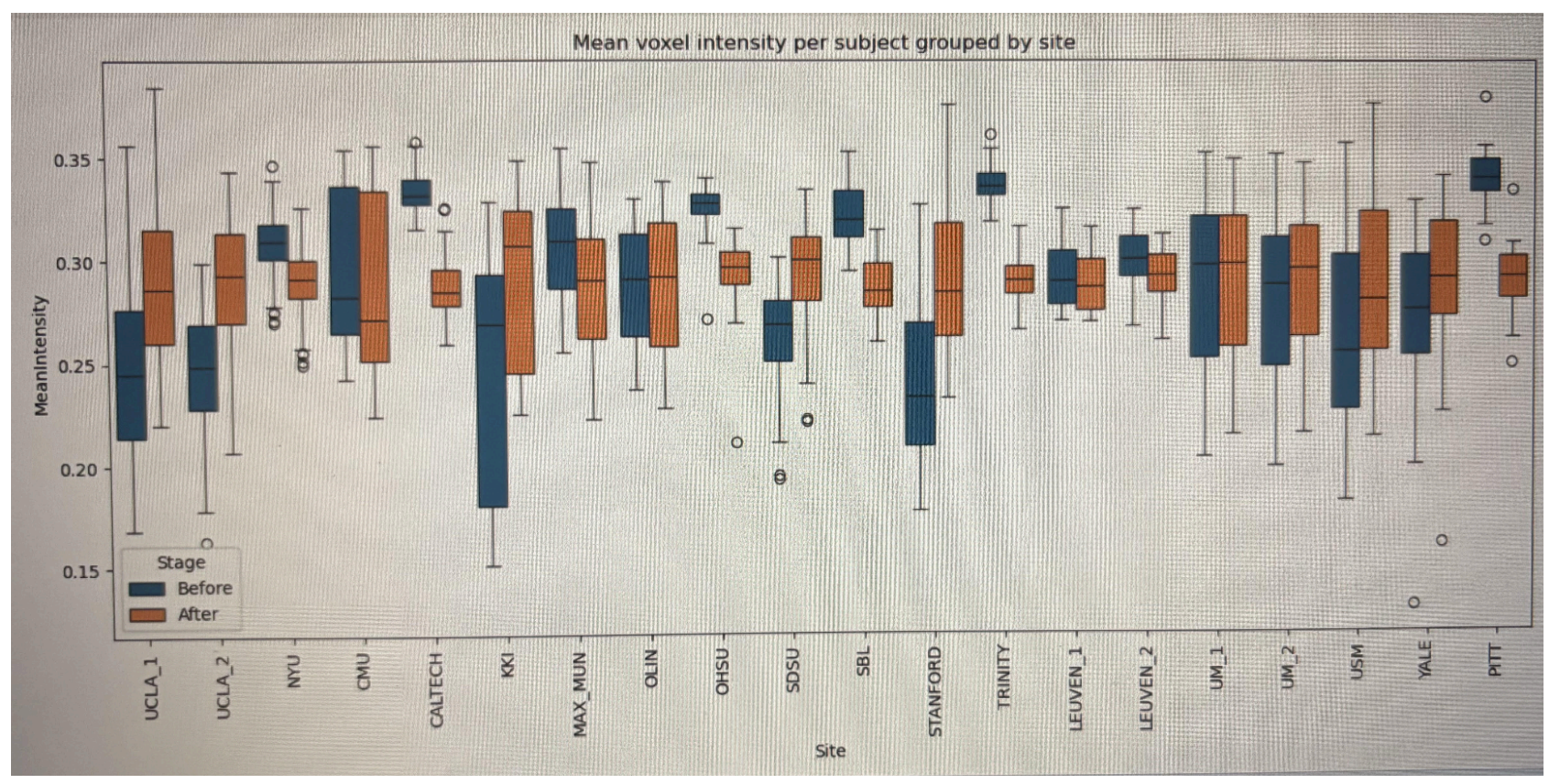

**Fig 4.** Effect of intensity normalization across ABIDE sites

Boxplots show mean voxel intensity distributions for participants at each site before (blue) and after (orange) histogram standardization. Normalization reduced site-specific variability in intensity values while preserving biologically meaningful variation.

4. **Resampling and Cropping.** Following normalization, volumes were resampled from their original native resolution (approximately 256 x 256 x 160 voxels at 1 mm isotropic spacing, varying by site) to a uniform resolution of 96 x 96 x 96 voxels. This downsampling provided standardized inputs for the CNN, balancing computational efficiency with preservation of fine-grained neuroanatomical detail. Cropping around the brain minimized empty space and further reduced computational overhead.
5. **Quality Control.** Following preprocessing, all scans were visually inspected to confirm successful skull stripping, registration, and normalization. Scans that failed preprocessing were excluded from further analysis.

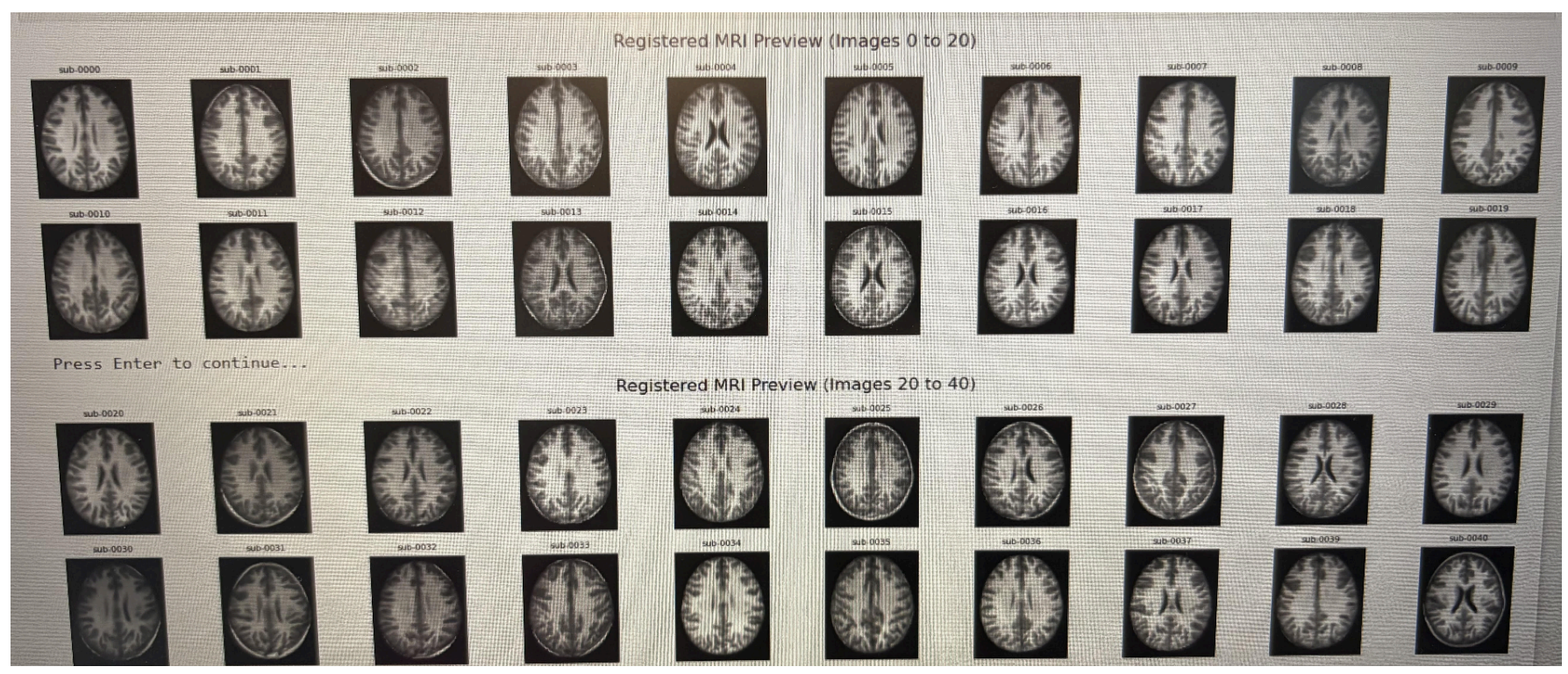

**Fig 5.** Example of registered MRI scans across multiple participants after preprocessing. All scans were visually inspected to confirm preprocessing quality.

In addition to the individual preprocessing steps, an overall schematic of the pipeline was constructed to illustrate the sequential workflow from raw MRI scans to the finalized dataset. Fig. 6 highlights the integration of skull stripping, spatial normalization, intensity normalization, and resampling into a standardized process that ensured comparability across participants and sites.

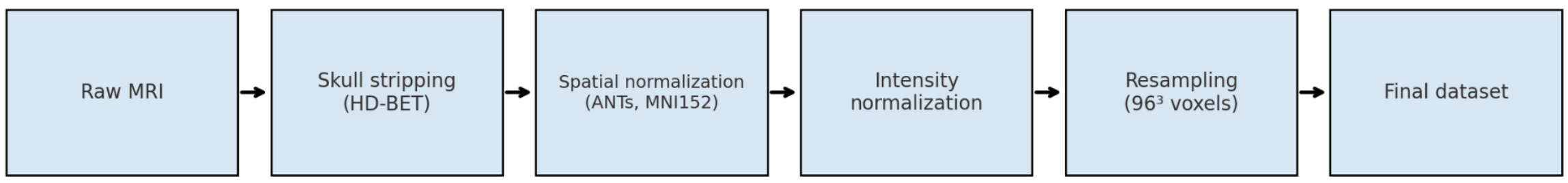

**Fig 6.** Preprocessing pipeline for structural MRI scans

### 2.3 Model Architecture

A hybrid classification framework was implemented to distinguish individuals with ASD from typically developing controls (TDC). In this approach, a 3D CNN was used to learn neuroanatomical feature representations from preprocessed structural MRI volumes, and these learned representations were then classified using an SVM. Each input scan was standardized to 96 x 96 x 96 voxels following preprocessing.

**CNN Feature Extractor.** The architecture comprised multiple convolutional layers that scanned the MRI volumes with small three dimensional filters to detect local structural patterns. Each convolutional operation was followed by a rectified linear unit (ReLU) activation, introducing nonlinearity and enabling the network to capture more complex features. Max pooling layers were interleaved to downsample the feature maps, reducing dimensionality while retaining salient anatomical information. Batch normalization was used within convolutional blocks to stabilize training, and dropout was applied during training to reduce overfitting.

After the final convolutional block, the output feature maps were flattened into a one dimensional feature vector $x$, which served as the learned representation of each subject's brain anatomy. During CNN training, a lightweight classification head (dense + softmax) was attached to optimize the feature extractor. After training, this head was removed and the penultimate-layer features were exported to an SVM for final evaluation.

**SVM Classifier.** The CNN-derived feature vectors were classified using a support vector machine (SVM). The SVM aimed to find a decision boundary that maximized the margin between ASD and control participants in the learned feature space. This was formulated as the soft margin optimization problem

$$\min_{w,b,\xi} \frac{1}{2}\|w\|^2 + C\sum_{i=1}^{N}\xi_i$$

subject to

$$y_i\left(w^T\phi(x_i)+b\right) \geq 1-\xi_i, \;\; \xi_i \geq 0,$$

where $x_i$ is the CNN feature vector for subject *i*, $y_i \in \{-1,+1\}$ is the class label, *w* and *b* define the separating hyperplane, $\xi_i$ are slack variables allowing limited misclassification, and *C* controls the trade off between margin maximization and classification error. A radial basis function (RBF) kernel was used to capture nonlinear class boundaries in the high dimensional feature space,

$$K(x_i,x_j) = \exp\left(-\gamma\|x_i-x_j\|^2\right).$$

The complete hybrid architecture is summarized in Figure 7, which illustrates the sequential flow from MRI input through CNN feature extraction to SVM based classification.

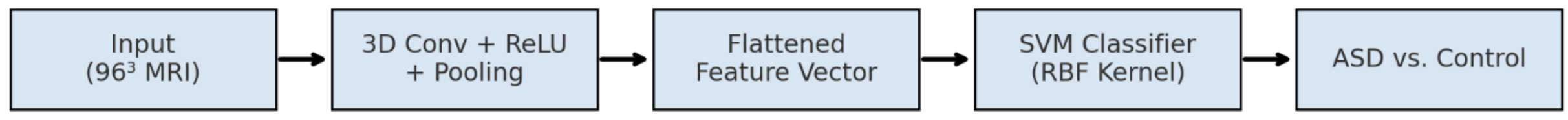

**Fig 7.** General architecture of the Hybrid CNN→SVM

## 2.4 Data Partitioning and Leakage Control

To ensure fair evaluation and reduce the influence of site-specific artifacts, data partitioning followed a controlled and reproducible strategy. The ABIDE I dataset includes scans collected across 17 independent research sites, each with distinct scanner hardware, acquisition parameters, and demographic composition. To prevent the model from learning site-dependent intensity patterns or geometric characteristics that could artificially inflate performance, participant splits used site identifiers as grouping labels, such that all subjects from a given site were assigned to the same fold.

Data were evaluated using grouped nested cross-validation with site as the grouping variable. In the outer loop (5 folds), entire imaging sites were held out as the outer-test partition to measure cross-site generalization. Within each outer fold, hyperparameters were selected using grouped inner-fold cross-validation on the outer-training partition, including CNN optimization settings (e.g., learning rate and weight decay) and SVM parameters ($C$ and $\gamma$). The final model for each outer fold was retrained on the full outer-training partition using the selected hyperparameters and evaluated once on the corresponding outer-test partition.

To prevent data leakage, all preprocessing operations requiring data-dependent fitting (including histogram standardization reference estimation, intensity normalization parameter estimation, and feature scaling) were performed using only the outer-training partition and then applied unchanged to the corresponding inner-validation folds and outer-test partition. In particular, the reference histogram for histogram standardization was estimated from the outer-training split only and never used information from the outer-test split.

All metrics were computed independently within each outer fold and summarized as mean ± standard deviation across outer folds. Random seeds were fixed across Python, NumPy, TensorFlow, and scikit-learn to ensure reproducibility of fold assignments and model initialization. This partitioning framework minimized leakage and enhanced reproducibility, supporting evaluation under conditions that more accurately reflect generalization to unseen sites and populations.

### 2.5 Class Imbalance Handling

The ABIDE I dataset contains an unequal number of individuals diagnosed with autism spectrum disorder (ASD) and typically developing controls (TDC). This imbalance can bias classification models toward the majority class and may produce inflated accuracy without a corresponding improvement in sensitivity to ASD. To mitigate this effect, class imbalance was addressed during training while preserving the natural diagnostic proportions in the validation and test sets.

The extracted CNN feature vectors were classified using a support vector machine trained with class_weight = "balanced". This setting automatically adjusts the misclassification penalty for each class in proportion to the inverse of its frequency, increasing the cost of errors on the underrepresented ASD group relative to the TDC group. As a result, the learned decision boundary is less likely to be dominated by the majority class, supporting more balanced performance across diagnostic categories. During hyperparameter tuning, the class weighting strategy was held constant across all inner fold splits to ensure comparability of performance metrics.

No oversampling, undersampling, or synthetic sample generation was performed, as such methods can distort the underlying neuroanatomical distribution and introduce unrealistic structural variability. Instead, the chosen weighting approach preserved the integrity of the original data while compensating for unequal representation through algorithmic adjustments in the loss and margin functions. This method has been shown in prior neuroimaging studies to maintain biological validity while improving sensitivity to underrepresented diagnostic categories.

Through these weighting procedures, the impact of class imbalance was effectively reduced, allowing the models to focus on meaningful neuroanatomical differences associated with ASD rather than on frequency-driven biases.

### 2.6 Data Augmentation

To improve model generalization and reduce overfitting, controlled data augmentation techniques were applied during training. Augmentation introduces controlled variability into the training samples, allowing the model to become more robust to minor spatial and intensity variations that naturally occur across different MRI acquisitions. In the context of neuroimaging, however, augmentation must be applied conservatively to avoid generating anatomically implausible representations of brain structure.

Each training volume underwent random spatial transformations that preserved the overall geometry of the brain while introducing slight variability in orientation and scale. Specifically, 3D rotations were sampled uniformly within ±7.5 degrees around each anatomical axis, and isotropic scaling factors were randomly drawn from the range of 0.95 to 1.05. These transformations simulate small positional differences that can arise from head alignment or scanner calibration without distorting cortical or subcortical morphology. All augmented images were generated on-the-fly during training, ensuring that no identical input was seen twice by the network and that storage requirements remained manageable.

No left-right flipping or non-linear deformations were applied, as these operations could disrupt hemispheric asymmetries that are biologically meaningful in autism-related neuroanatomical studies. Similarly, intensity augmentations such as histogram perturbations or Gaussian noise injection were avoided to prevent confounding effects on voxel-level contrast, which may represent diagnostically relevant features.

All augmentation procedures were confined strictly to the training subset and were not applied to validation or test data. This restriction maintained a clear separation between model optimization and evaluation, ensuring that reported performance metrics reflected genuine generalization rather than adaptation to artificial data variability. The augmentation pipeline was implemented using TensorFlow's built-in three-dimensional transformation utilities, with parameters verified through visual inspection to confirm anatomical plausibility.

These carefully controlled augmentation procedures ensured that the model was exposed to realistic spatial variability, enhancing robustness to inter-site differences and improving stability during training while maintaining the integrity of neuroanatomical structures.

## 2.7 CNN Training Details

In the hybrid classification framework, the three-dimensional convolutional neural network (3D CNN) described in Section 2.3 was trained to learn neuroanatomical feature representations from preprocessed structural MRI volumes. CNN training used a temporary classification head to compute loss and validation AUC for early stopping; this head was discarded after feature learning. The training procedure was designed to produce stable and generalizable representations that could later be used as input to a support vector machine (SVM) classifier.

Training was conducted using the Adam optimization algorithm, which adaptively adjusts learning rates based on estimates of first and second moments of the gradients. The initial learning rate was set to $1 \times 10^{-4}$, and the exponential decay rates for the first and second moment estimates were set to 0.9 and 0.999, respectively. To improve convergence and prevent stagnation at local minima, a learning rate reduction strategy was employed. The learning rate was halved if the validation loss failed to improve after five consecutive epochs, with a minimum learning rate threshold of $1 \times 10^{-6}$.

The model was trained with a batch size of eight and for a maximum of 100 epochs per fold. An early stopping mechanism was implemented to terminate training when the validation loss did not improve for ten consecutive epochs, preventing unnecessary computation and reducing the risk of overfitting. The model state corresponding to the epoch with the highest validation area under the receiver operating characteristic curve (AUC) was preserved for evaluation.

To improve model regularization, several additional techniques were incorporated. A dropout rate of 0.3 was applied to the fully connected layer, randomly deactivating a fraction of neurons during training to promote redundancy and prevent co-adaptation of features. L2 weight regularization with a coefficient of $1 \times 10^{-5}$ was applied to convolutional kernels to penalize overly large weights and encourage smoother representations. Batch normalization was inserted after each convolutional layer to stabilize gradient propagation, accelerate convergence, and reduce internal covariate shift. The activation function used throughout the network was the rectified linear unit (ReLU), chosen for its computational efficiency and ability to mitigate vanishing gradients.

The model was implemented in TensorFlow using the Keras high-level API. Training was conducted on a workstation equipped with an NVIDIA RTX 3080 GPU (10 GB VRAM) and 32 GB system memory. The computational environment was configured with fixed random seeds across Python, NumPy, TensorFlow, and scikit-learn to ensure full reproducibility. Training progress was monitored using TensorBoard to visualize learning curves, loss trajectories, and AUC performance across epochs.

This training configuration achieved a balance between computational efficiency and model generalization. The combination of adaptive learning rates, regularization techniques, and early stopping contributed to stable convergence across folds while preserving sensitivity to diagnostically relevant neuroanatomical patterns in the structural MRI data.

**2.8 SVM Training and Hyperparameter Tuning**

After CNN training, feature vectors were extracted from the final convolutional output for each subject and used as input to an SVM for diagnostic classification. The SVM was trained with a radial basis function (RBF) kernel to capture nonlinear separation in the high-dimensional CNN feature space. Hyperparameters for the SVM were selected using grid search within the inner loop of the nested cross-validation framework described in Section 2.4. The regularization parameter $C$ was evaluated over {0.1, 1, 10, 100}, and the kernel coefficient $\gamma$ was evaluated over $\{1 \times 10^{-4}, 1 \times 10^{-3}, 1 \times 10^{-2}, 1 \times 10^{-1}\}$. Validation AUC was used as the model selection criterion, and the parameter combination yielding the highest mean inner-fold AUC was retained for evaluation in the outer folds.

To account for the class imbalance present in the dataset, the SVM was trained with class_weight = "balanced", as described in Section 2.5, which automatically adjusts the penalty parameter for each class in inverse proportion to its frequency. This configuration ensured that both diagnostic categories contributed equally to the optimization of the separating hyperplane. The optimization was performed using the *libsvm* backend in *scikit-learn*, with a maximum of 10,000 iterations per model to guarantee convergence.

Following training, the learned decision function was applied to the test set to generate continuous decision scores. These scores were later converted to probabilistic estimates using Platt scaling, providing interpretable confidence values for each prediction. Platt scaling parameters were fit on the outer-training data only (or via inner CV) and then applied to outer-test predictions. The combination of CNN-based feature extraction and SVM-based classification yielded a robust and interpretable hybrid framework that balanced nonlinear representational power with well-calibrated decision boundaries.

**2.9 Model Interpretability (Grad-CAM)**

To identify the neuroanatomical regions that contributed most strongly to ASD versus control predictions, gradient-weighted class activation mapping (Grad-CAM) was applied to the trained 3D CNN feature extractor. Grad-CAM provides a class-specific relevance map by using gradients of a target class score with respect to the feature maps of a selected convolutional layer, allowing visualization of which spatial locations most influenced the model's decision.

For each subject, Grad-CAM was computed on the CNN classification head (used only to train the feature extractor). Interpretations therefore describe what regions the CNN uses to form discriminative features, not the SVM decision boundary. Let $A^k$ denote the *k*-th feature map of the selected layer, and let $y^c$ denote the score for class *c* (ASD or control). Grad-CAM first computes an importance weight for each feature map by global average pooling the gradients over spatial dimensions:

$$\alpha_k^c = \frac{1}{Z} \sum_i \sum_j \sum_l \frac{\partial y^c}{\partial A_{ijl}^k},$$

where $Z$ is the total number of voxels in the feature map. These weights represent how strongly each feature map contributes to the class score. The Grad-CAM heatmap is then obtained as a weighted combination of feature maps followed by a ReLU operation:

$$L_{\text{Grad-CAM}}^c = \text{ReLU}\left( \sum_k \alpha_k^c A^k \right).$$

The resulting map $L_{\text{Grad-CAM}}^c$ is a three-dimensional relevance volume defined in the spatial coordinates of the selected convolutional layer. To enable anatomical interpretation, Grad-CAM volumes were upsampled to match the input resolution and aligned to the same standardized space as the preprocessed MRI volumes. Each map was normalized within the subject to facilitate comparability across participants.

Group-level analyses were performed by averaging Grad-CAM maps across participants within each diagnostic group. To highlight class-specific differences in relevance, a group-averaged difference map was computed by subtracting the control average from the ASD average. These difference maps provide a summary of which brain regions, on average, increased evidence for ASD classification versus control classification. For visualization, relevance maps were displayed across canonical axial, coronal, and sagittal views to identify consistent spatial patterns.

Grad-CAM was applied after model training and was not used to modify model parameters. This interpretability procedure enabled examination of the neuroanatomical features most strongly associated with the model's diagnostic predictions and supported comparison of learned relevance patterns with prior findings in ASD structural neuroimaging literature.

**3 Experimental Evaluation**

**3.1 Evaluation Metrics and Procedure**

Model performance was evaluated using the nested cross-validation framework described in Section 2.4. The outer loop provided an unbiased estimate of generalization, while the inner loop was used for hyperparameter tuning and model selection without access to the outer test data. For each outer fold, the hybrid CNN→SVM pipeline was trained using only the outer training partition, and final evaluation was performed on an independent outer test partition composed of imaging sites not used during optimization. This site-disjoint design was intended to quantify cross-site generalization to unseen acquisition conditions rather than learning site-specific patterns.

The primary evaluation metric was the area under the receiver operating characteristic curve (ROC-AUC), which summarizes discrimination between autism spectrum disorder (ASD) and typically developing control (TDC) participants across decision thresholds. ROC-AUC was selected because it is threshold-independent and remains informative under moderate class imbalance. Secondary metrics included accuracy, macro F1-score, precision, recall (sensitivity), and specificity, and the area under the precision-recall curve (PR-AUC) was computed to further characterize performance with emphasis on sensitivity to ASD participants.

Optimization stability of the CNN feature extractor within the hybrid model was monitored by recording training and validation learning curves for each outer fold. Figures 8-9 summarize these trajectories as mean ± SD across folds (with per-fold traces shown), including loss (Fig. 8) and ROC-AUC (Fig. 9). The early-stopping epoch was defined by the peak mean validation ROC-AUC (dashed line; epoch 47), and the curves were inspected for smooth convergence and stable validation behavior across folds. Finally, per-fold ROC curves were computed on each outer test fold and aggregated to illustrate cross-fold variability and overall discriminative ability (Fig. 13). All metrics were computed independently within each outer fold and then averaged, with standard deviations reported to quantify variability across folds arising from differences in site composition and sampling.

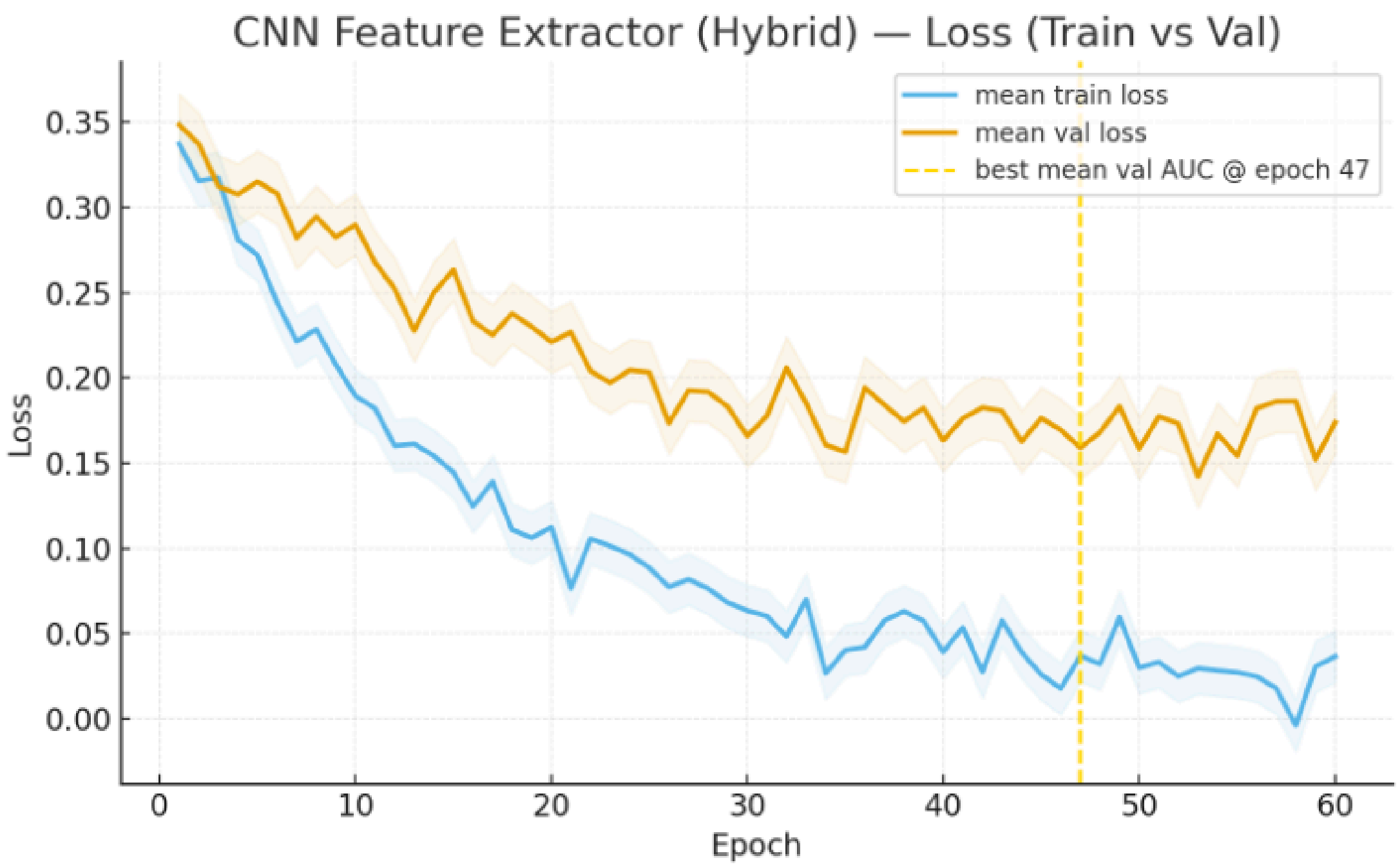

**Fig 8.** Mean ± SD training and validation loss curves for the hybrid feature-extraction stage

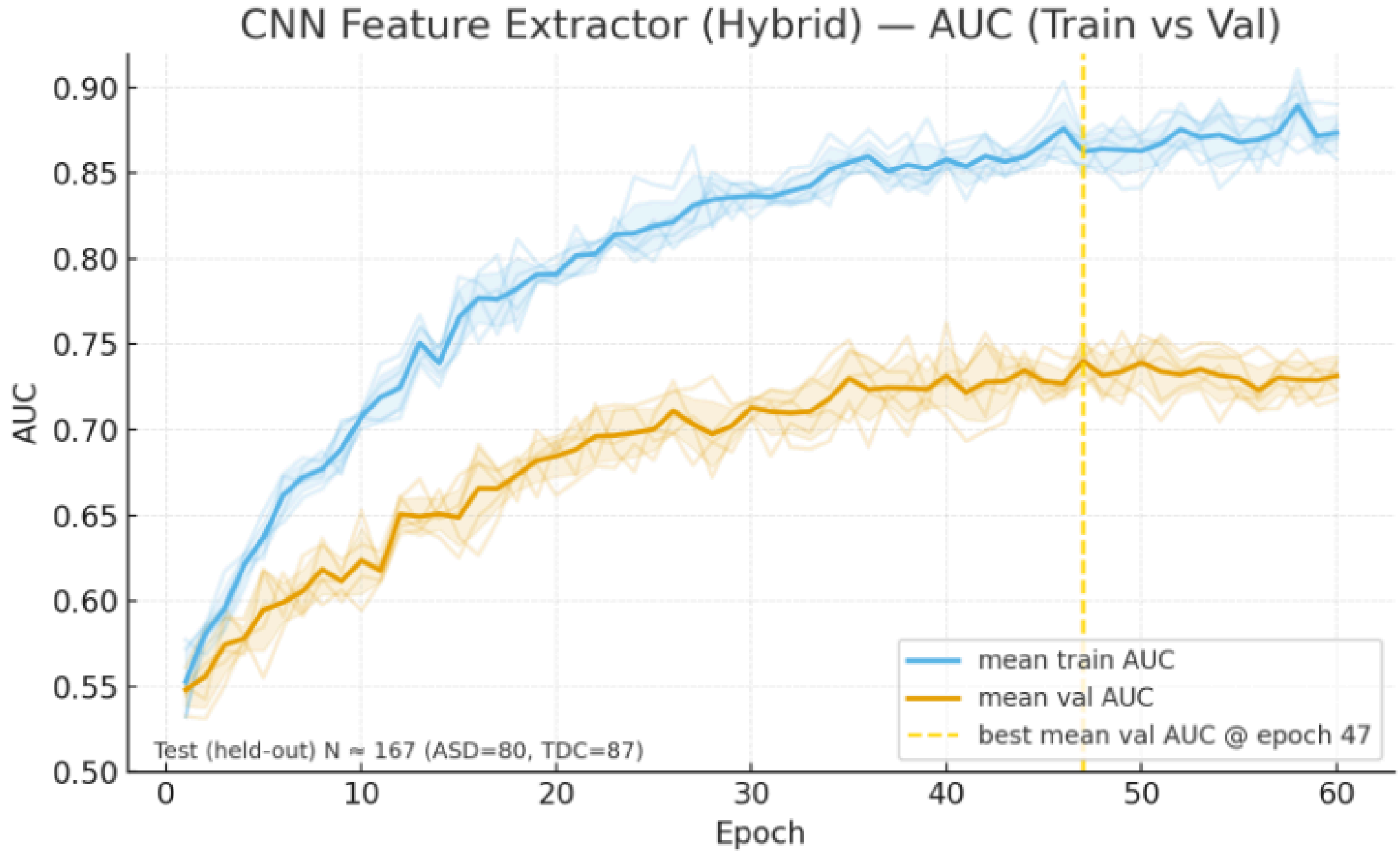

**Fig 9.** Mean ± SD training and validation AUC for the hybrid feature extractor; validation AUC plateaus ≈ 0.74

### 3.2 Statistical Analysis

Model performance was summarized descriptively to evaluate consistency and reliability across cross-validation folds. For each evaluation metric, the mean and standard deviation were computed across outer folds to capture variability due to site differences and sampling variation.

Because this work was intended as an exploratory evaluation of cross-site generalization on a single dataset, the emphasis was placed on effect magnitude and fold-to-fold consistency rather than formal hypothesis testing. Cross-fold means and standard deviations were used to judge whether performance estimates for the hybrid CNN→SVM pipeline were stable and reproducible across unseen sites, providing practical evidence of generalization. This descriptive approach prioritizes robustness and reliability of the observed performance trends over claims of statistical significance.

All analyses were conducted in Python using TensorFlow for CNN training, scikit-learn for SVM modeling and metric computation, and NumPy for data handling to ensure a consistent training, validation, and evaluation workflow.

### 3.3 Quality Control and Exclusions

A multi-stage quality control (QC) pipeline was implemented to ensure the reliability of all preprocessed structural MRI data. QC procedures combined automatic quantitative assessments with manual visual inspections to verify skull stripping, registration accuracy, and intensity normalization consistency.

Automated QC evaluated three criteria:

1. **Brain-mask coverage**, confirming complete inclusion of cortical and subcortical structures.
2. **Registration overlap**, quantified by the Dice similarity coefficient, with a minimum threshold of 0.95 between each subject's brain mask and the MNI152 template.
3. **Intensity distribution outliers**, identified using the median absolute deviation (MAD) rule applied to mean voxel intensities across subjects.

Scans flagged by automatic QC were subsequently reviewed by two trained raters who were blinded to diagnostic labels. Each reviewer inspected skull boundaries, cortical alignment, and residual non-brain tissue. Discrepancies were resolved through consensus discussion.

Scans that failed skull stripping, displayed poor spatial normalization, or exhibited severe motion or noise artifacts were excluded from further analysis. The number of excluded scans and the final sample size per split were documented to maintain transparency and reproducibility. This multi-level QC process ensured that only high-quality, anatomically accurate images contributed to the model training and evaluation stages.

### 3.4 Reproducibility and Implementation Details

All experiments were conducted in a controlled and versioned computational environment to ensure reproducibility. Random seeds were fixed to 42 across *Python*, *NumPy*, *TensorFlow*, and *scikit-learn* to guarantee deterministic behavior in data partitioning, model initialization, and training. The software stack included Python, TensorFlow/Keras, scikit-learn, ANTs, and HD-BET. Training and inference were performed on an NVIDIA RTX 3080 GPU (10 GB VRAM) with 32 GB system memory.

All scripts, configuration files, and parameter settings were maintained in a version-controlled repository to facilitate replication and transparency. Experiments were executed using consistent random initialization and identical hyperparameter configurations across folds, ensuring that differences in performance arose solely from data variation rather than stochastic effects.

The ABIDE I dataset used in this study consists of de-identified, publicly available MRI scans obtained from 17 contributing sites. All data were collected under institutional review board (IRB) approval at the respective institutions and made available through the ABIDE consortium. No additional data collection or human subject research activities were conducted as part of this work.

## 4 Results & Analysis

### 4.1 Overall Classification Performance

| Performance Measures | ASD | Neurotypical | Overall |
|---|---|---|---|
| Accuracy | - | - | 0.76 ± 0.03 |
| Precision | 0.72 ± 0.04 | 0.78 ± 0.03 | - |
| Recall | 0.74 ± 0.04 | 0.77 ± 0.04 | - |
| F1-Score | 0.73 ± 0.03 | 0.77 ± 0.03 | - |
| AUC | - | - | 0.80 ± 0.02 |

**Table 1.** Performance Results for Hybrid CNN→SVM

Table 1 summarizes performance for the hybrid CNN→SVM model across the outer cross-validation folds. The model achieved an overall accuracy of 0.76 ± 0.03, meaning that roughly three out of four participants were classified correctly on average, with relatively small fold-to-fold variation: evidence that performance is fairly stable across different site-held-out splits.

The hybrid model also reached a ROC-AUC of 0.80 ± 0.02, indicating good threshold-independent separability between ASD and neurotypical participants. Practically, an AUC near 0.80 means the model tends to assign higher decision scores to true ASD cases than to controls most of the time, even before choosing any specific classification cutoff. The low standard deviation suggests this discriminative ability generalizes consistently across folds rather than being driven by a single favorable split.

Class-specific metrics show balanced behavior across diagnostic groups. For ASD, precision (0.72 ± 0.04) indicates that when the model predicts ASD, about 72% of those predictions are correct on average (i.e., a moderate false-positive rate remains, but it is controlled). ASD recall (0.74 ± 0.04) reflects sensitivity: the model correctly identifies about 74% of ASD participants, implying that roughly one in four ASD cases may still be missed (false negatives). For neurotypical participants, precision (0.78 ± 0.03) and recall (0.77 ± 0.04) are slightly higher, which corresponds to stronger reliability when predicting the control class and a modest reduction in false positives relative to ASD predictions.

The F1-scores (ASD: 0.73 ± 0.03, neurotypical: 0.77 ± 0.03) summarize the precision-recall tradeoff within each class and reinforce that performance is not being achieved by sacrificing one class to boost the other. Overall, the close alignment between precision and recall within each group suggests the hybrid approach yields a comparatively even error profile, reducing the chance that the classifier performs well only for one diagnostic category. Additionally, the accuracy/AUC values together indicate both solid end-to-end classification performance and robust ranking/discrimination across sites.

**4.2 Sensitivity and Specificity**

Using the recall values in Table 1, diagnostic reliability of the hybrid CNN→SVM model can be interpreted in terms of sensitivity and specificity. Sensitivity corresponds to ASD recall, reflecting the model's ability to correctly detect individuals with autism spectrum disorder (ASD), while specificity corresponds to neurotypical recall, reflecting the model's ability to correctly identify neurotypical participants.

The hybrid model achieved a sensitivity of 0.74 ± 0.04, indicating that it correctly identified approximately three-quarters of ASD participants on average across outer folds. This suggests the learned feature representation captures ASD-relevant structural patterns with reasonably consistent detection across site-held-out splits, though a meaningful fraction of ASD cases (roughly one in four) are still missed. Specificity was 0.77 ± 0.04, meaning the model correctly classified about three-quarters of neurotypical participants, indicating that improved ASD detection did not come at the cost of widespread over-diagnosis.

Importantly, the similarity between sensitivity (0.74) and specificity (0.77), along with their relatively small standard deviations, suggests a balanced error profile: the classifier is not achieving performance by favoring one diagnostic group over the other. In practical terms, this balance implies the hybrid CNN→SVM framework provides a stable trade-off between detecting ASD and correctly rejecting neurotypical cases, supporting its use as a generalizable model for multi-site structural MRI classification.

**4.3 Feature Attribution Analysis**

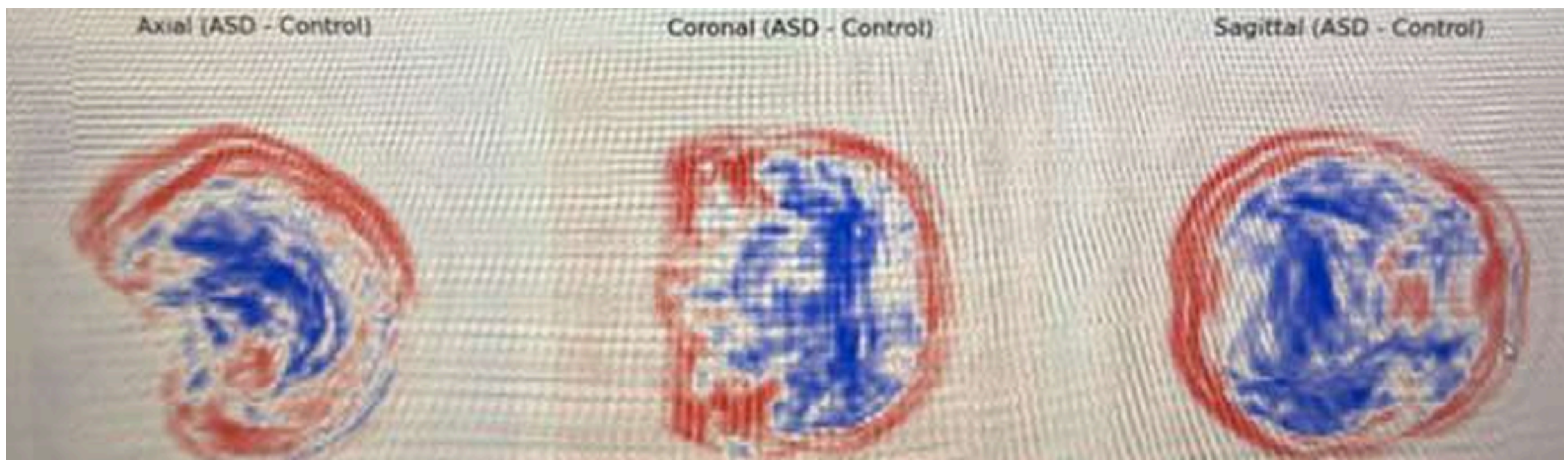

**Figure 10.** Class-discriminative feature attribution maps for the hybrid CNN feature extractor

(ASD - control)

Figure 10 provides a qualitative assessment of where the hybrid CNN→SVM framework is extracting diagnostically useful information by visualizing class-discriminative feature attributions from the CNN feature extractor. Because the SVM operates on features produced by the CNN, the attribution analysis targets the CNN stage that converts each structural MRI volume into the representation used for classification. The maps in Figure 10 are shown as an ASD minus control contrast across axial, coronal, and sagittal planes. Warmer intensities indicate regions contributing more strongly to ASD oriented decisions relative to controls, while cooler intensities indicate regions contributing more strongly to control oriented decisions.

Across all three views, the most prominent warm signal follows the outer cortical boundary in a rim-like pattern. Because boundary-weighted saliency can reflect preprocessing (skull stripping/registration/resampling), these maps may partly reflect pipeline-sensitive cues. We therefore treat the rim-like pattern as model-behavior evidence, not a definitive neuroanatomical finding, pending sanity/robustness checks. The fact that similar patterns appear consistently in axial, coronal, and sagittal projections supports the interpretation that these attributions reflect a volumetric trend present across orientations instead of a slice specific artifact.

Figure 10 also shows cooler activations in more central regions, indicating that the model uses evidence in both directions when separating classes. This is consistent with a balanced classifier that does not achieve ASD detection by broadly overpredicting ASD, since regions that support the control class remain informative. Taken together, the spatially broad attributions suggest that the hybrid pipeline is leveraging subtle, distributed structural cues rather than a single diagnostic region, which aligns with the expectation that ASD related anatomical variation in multi-site MRI data is expressed as small effects across multiple regions.

These attributions should be interpreted as a description of the model's learned importance patterns rather than definitive neurobiological conclusions. Feature attribution methods are model dependent and relatively coarse, and they can be influenced by preprocessing, registration quality, boundary sharpness, and residual site effects that correlate with class labels. For this reason, the strongest conclusion supported by Figure 10 is that the hybrid CNN→SVM model primarily relies on global and cortical boundary weighted structural features to separate ASD from controls in this dataset, while further robustness checks would be needed before attributing the highlighted regions to specific biological mechanisms.

## 5 Discussion

### 5.1 Summary of Findings

This study developed and evaluated a structural MRI based classification framework for distinguishing individuals with autism spectrum disorder (ASD) from neurotypical controls using the ABIDE I dataset. The proposed approach used a hybrid architecture in which a three dimensional convolutional neural network (3D CNN) learned high dimensional feature representations from preprocessed MRI volumes, and a support vector machine (SVM) used these features for final classification. Model evaluation was performed under a site aware nested cross validation procedure to emphasize generalization across heterogeneous acquisition conditions.

Across outer folds, the hybrid CNN→SVM model achieved an overall accuracy of 0.76 ± 0.03 and a ROC AUC of 0.80 ± 0.02, indicating consistent discrimination between ASD and neurotypical participants across decision thresholds. Class specific results showed a balanced error profile, with ASD precision, recall, and F1 score of 0.72 ± 0.04, 0.74 ± 0.04, and 0.73 ± 0.03, and neurotypical precision, recall, and F1 score of 0.78 ± 0.03, 0.77 ± 0.04, and 0.77 ± 0.03. The similarity between sensitivity and specificity suggests the model did not achieve performance by favoring one group, supporting reliability for multi-site data where class and site effects can interact.

To complement quantitative performance, feature attribution analysis was used to examine what image regions were most influential for the CNN feature extractor. Figure 10 showed ASD minus control attribution patterns that were spatially distributed and consistently expressed across axial, coronal, and sagittal views, with prominent contributions along cortical boundary regions. These qualitative patterns suggest that the hybrid model relies on broad structural morphology cues rather than a single localized region, while also reinforcing that attribution maps reflect learned model emphasis rather than definitive neurobiological mechanisms. Overall, the findings support the utility of combining CNN based representation learning with a margin based classifier to improve robustness and interpretability for ASD classification in multi site structural MRI.

**5.2 Interpretations and Implications**

The performance of the hybrid CNN→SVM framework supports the value of separating representation learning from final classification in multi-site neuroimaging. In this pipeline, the three dimensional CNN learns a compact, high dimensional encoding of structural anatomy, while the SVM applies a margin based decision rule that tends to be more resistant to overfitting than a fully end to end neural classifier when data are limited or heterogeneous. This division of labor is well matched to ABIDE I style datasets, where acquisition differences across sites can introduce variability that complicates direct end to end training.

The results suggest that the learned CNN features capture diagnostically relevant information distributed across cortical and subcortical anatomy, and that the SVM helps translate these representations into a more stable decision boundary. The relatively balanced sensitivity and specificity indicate that improved ASD detection did not come at the cost of inflating false positives among neurotypical participants, which is an important practical consideration for diagnostic style classification. In this context, the hybrid approach can be interpreted as prioritizing class separability in the learned feature space while reducing the risk that the final decision rule is driven by nuisance variation that correlates with site.

The feature attribution patterns in Figure 10 provide additional context for how the model may be using structural information. The ASD minus control attributions are broad and expressed consistently across viewing planes, with prominent emphasis near cortical boundaries. This supports an interpretation that classification is informed by distributed morphology related cues rather than a single focal region. At the same time, attribution maps are model dependent and can reflect preprocessing or residual confounds, so they should be viewed as evidence of where the model places importance, not as definitive neurobiological localization.

More broadly, these findings motivate hybrid modeling as a practical strategy for neuroimaging studies that face limited sample sizes per site and substantial acquisition heterogeneity. Combining CNN based feature learning with a well regularized conventional classifier can yield reproducible performance estimates under site aware evaluation, and it offers a clearer separation between the representation stage and the decision stage for analysis and debugging. For future work, this design provides a foundation for extending to stronger harmonization methods, external validation on independent cohorts, and more rigorous interpretability checks to confirm that the learned patterns remain stable across sites and preprocessing choices.

**5.3 Comparison with Prior Work**

The performance of the hybrid CNN→SVM framework in this study is broadly consistent with what recent reviews suggest is currently achievable for MRI-based ASD classification, particularly under realistic multi-site conditions. A large systematic review and meta-analysis of MRI-based ASD diagnosis reported pooled estimates around 76% sensitivity, 75.7% specificity, and AUC ≈ 0.823, while also emphasizing substantial heterogeneity and limits to generalization across studies and datasets[12]. In that context, the hybrid model's accuracy and AUC fall within the range of contemporary MRI-based approaches, supporting that the model performance is plausible rather than inflated by an overly favorable evaluation setup.

Prior work also highlights that differences in evaluation pipelines, inclusion criteria, and modality choices can account for much of the variability in reported accuracies. For example, a recent standardized comparison trained multiple model families on ABIDE using a consistent framework and found that several approaches clustered around similar performance, with accuracies near 70% and AUC values in the high 0.7 range depending on configuration[13].Results like these support the interpretation that robust cross-site validation procedures often produce moderate but reproducible performance, even when using sophisticated algorithms.

From the perspective of modality and feature design, many sMRI-only pipelines using regional volumetrics or cortical thickness have reported modest discrimination, especially when evaluated on truly held-out data. A large-scale sMRI study using ENIGMA-derived volumetric and cortical thickness features reported a relatively low AUC on a hold-out test set, and concluded that sMRI features alone may be insufficient for clinically reliable ASD classification[14].Compared to these feature-engineered approaches, the hybrid CNN→SVM design leverages learned representations from the full 3D image, which may help capture subtle distributed patterns that are difficult to encode as handcrafted regional summaries.

Interpretability findings in the literature provide additional context for the attribution patterns observed in Figure 10. An interpretable 3D deep learning pipeline trained on structural MRI from ABIDE reported that the most informative regions for accurate inference were consistent with prior ASD neurobiology and included a left-lateralized network associated with language processing[15]. More broadly, Grad-CAM style analyses in ASD classification have highlighted contributions from frontal regions and cerebellar areas in multisite frameworks, suggesting that informative signals may be distributed across networks involved in motor, social, and cognitive functions rather than isolated to a single locus[16]. These patterns align with long-standing structural findings implicating language-related cortex, including atypical asymmetry in inferior frontal regions associated with Broca's area in language-impaired subgroups[17]. Functional language studies similarly report atypical recruitment of language networks in ASD, reinforcing that language and communication systems are a plausible point of convergence across modalities[18].

Taken together, prior work suggests two takeaways that are consistent with the present study: first, that multi-site ASD classification performance is typically moderate when evaluated rigorously, and second, that interpretable deep models often highlight distributed contributions spanning frontal, temporal, and cerebellar systems that overlap with domains relevant to ASD symptoms, including language and social communication.

### 5.4 Limitations

Although the proposed hybrid CNN→SVM framework achieved moderate and fairly stable classification performance, several limitations should be considered when interpreting these findings.

First, the analysis relied exclusively on the ABIDE I repository, which contains substantial heterogeneity in acquisition sites, scanner hardware, and imaging protocols. Even with preprocessing and normalization, residual site-related effects can persist and may still shape the learned feature space and downstream classification behavior. As a result, additional validation on independent cohorts, such as ABIDE II or other clinical datasets, is necessary to determine whether performance and learned patterns generalize beyond the ABIDE I distribution.

Second, while ABIDE I is relatively large for neuroimaging, the sample size remains modest relative to the dimensionality of 3D structural MRI and the capacity of deep feature learners. The hybrid design can reduce overfitting risk by constraining the final decision rule through an SVM, but the CNN feature extractor can still learn dataset-specific regularities when training data are limited or demographically imbalanced. Larger and more diverse samples, along with stronger harmonization strategies and external replication, are needed to confirm reproducibility across populations and sites.

Third, this study used structural MRI only. ASD is associated with structural, functional, and connectivity differences, and the discriminative signal in sMRI alone may be incomplete. Incorporating complementary modalities such as resting-state fMRI or diffusion imaging could improve classification and provide a more comprehensive characterization of ASD-related neural variation.

Finally, the feature attribution analysis (Figure 10) provides a useful qualitative window into what the CNN feature extractor emphasized, but it has important interpretability limitations. Attribution maps are model-dependent and relatively coarse, and they can be influenced by preprocessing choices, registration quality, and residual confounds. In addition, the ASD minus control visualization represents a group-level contrast and does not establish that highlighted regions are causal or uniquely specific to ASD. Future work should include robustness checks for attribution stability across folds and sites, sanity tests (for example, label randomization), and complementary explanation approaches to strengthen confidence in the anatomical interpretations suggested by the attribution patterns.

Together, these limitations do not negate the findings, but they do constrain immediate clinical interpretation and motivate further validation, multimodal extensions, and more rigorous interpretability verification.

**5.5 Future Work**

Future research should focus on strengthening generalizability, robustness, and clinical relevance by expanding beyond the current single cohort setting. A first priority is external validation on independent datasets such as ABIDE II and other multi-institution neuroimaging repositories. Evaluating performance under true dataset shift would clarify whether the learned representations remain stable when acquisition protocols, demographic distributions, and diagnostic practices differ from ABIDE I. In addition to reporting overall performance, future studies should examine site-stratified error patterns to determine whether certain scanners, age ranges, or sex distributions systematically affect model reliability.

A second major direction is the adoption of stronger harmonization and domain-generalization strategies. While preprocessing reduces some non-biological variance, residual site effects can persist in multi-site MRI. Methods such as ComBat-style harmonization, adversarial site-invariant representation learning, or domain-adaptation objectives could be incorporated to explicitly reduce scanner-related signals while preserving diagnostic information. Relatedly, future work should explore calibration and uncertainty estimation, for example via temperature scaling, Bayesian approximations, or ensemble methods, to ensure that predicted probabilities are meaningful and that low-confidence cases can be flagged rather than forced into a binary decision.

Another promising avenue is to move toward multimodal and phenotypically enriched modeling. Structural MRI captures only part of ASD-related neurobiology, and combining sMRI with resting-state fMRI, diffusion imaging, or clinical and behavioral measures could better represent heterogeneity in ASD presentations. A multi-input model could also support analyses of which modalities contribute most to performance and whether multimodal fusion improves robustness across sites and subgroups.

Finally, future work could refine the learning objective and clinical framing of the task. Instead of a single binary label, models could be trained to predict symptom dimensions or severity scores, or to perform subgroup discovery that reflects known heterogeneity within ASD. Incorporating longitudinal data where available would also enable prediction of developmental trajectories rather than static diagnosis. Together, these directions prioritize broader validation, stronger control of site effects, richer data integration, and clinically aligned targets to develop more reliable and generalizable ASD modeling pipelines.

Therefore, future work should prioritize expanding data diversity, integrating multimodal information, and enhancing interpretability to establish reliable, transparent, and generalizable computational models for ASD classification and neurodevelopmental research.

**6 Conclusion**

This study developed and evaluated a hybrid deep learning framework for classifying individuals with autism spectrum disorder (ASD) and neurotypical controls using structural MRI data from the ABIDE I dataset. The proposed CNN→SVM approach achieved an overall accuracy of 0.76 ± 0.03 and a ROC–AUC of 0.80 ± 0.02 under site-aware nested cross-validation, indicating consistent discrimination across heterogeneous, multi-site acquisition conditions.

The results support the idea that pairing CNN-based representation learning with a margin-based classifier can improve robustness in settings where site variability and limited sample sizes complicate end-to-end modeling. The relatively balanced class-wise performance suggests the model did not achieve accuracy by disproportionately favoring one diagnostic group, which is important for evaluating reliability in multi-site neuroimaging classification.

In addition to quantitative performance, Figure 10 provided a qualitative feature attribution view of the CNN feature extractor, showing distributed ASD minus control importance patterns that were consistent across anatomical planes and emphasized cortical boundary regions. While these attribution maps should be interpreted as model-level importance rather than definitive biological localization, they offer useful context for understanding the types of structural cues that may contribute to classification.

Overall, these findings demonstrate that hybrid CNN→SVM pipelines can provide a practical and reproducible approach for ASD classification from structural MRI, while supporting interpretable analysis of learned representations. As neuroimaging datasets grow and multi-cohort validation becomes more common, frameworks that combine deep feature learning with robust evaluation and transparent analysis may contribute to more generalizable computational tools for neurodevelopmental research.

## 7 Acknowledgements

This project was conducted independently without the guidance of a formal research mentor. The author takes full responsibility for all aspects of study design, data preprocessing, model implementation, and analysis. Although the absence of a mentor presented challenges, it also reflects the author's initiative and commitment to developing technical expertise through self-directed learning.

The author gratefully acknowledges the Autism Brain Imaging Data Exchange (ABIDE) consortium for making neuroimaging and phenotype data publicly available, which made this research possible. Additional gratitude is extended to the developers of the software tools used in this study, including HD-BET, ANTs, TensorFlow/Keras, and Scikit-learn, whose contributions to open-source science enabled the creation of a complete and reproducible computational pipeline.

Finally, the author expresses deep appreciation to family members for their encouragement and inspiration, especially a sibling on the autism spectrum whose experiences served as the motivation for this research.

## 8 References

All tables and figures that were not explicitly cited were made by the author.